\documentclass[reprint,aps,prl]{revtex4-1}
\usepackage{amssymb,amsmath,amsfonts}
\usepackage{graphicx}
\usepackage{xcolor}
\usepackage{hyperref}
\usepackage{braket}
\hypersetup{
	colorlinks,
	linkcolor={blue!60!black},
	citecolor={blue!60!black},
	urlcolor={blue!60!black}
}

\begin{document}

\title{
Probing Quantum Memory Effects with High Resolution
}

\author{Matthias Wittemer}
\author{Govinda Clos}
\author{Heinz-Peter Breuer}
\author{Ulrich Warring}
\author{Tobias Schaetz}

\affiliation{Physikalisches Institut, Universit\"at Freiburg, 
Hermann-Herder-Stra{\ss}e 3, D-79104 Freiburg, Germany}

\begin{abstract}
	Trapped atomic ions enable a precise quantification of the flow of information between internal and external degrees of freedom by employing a non-Markovianity measure [H.-P.~Breuer et al., Phys.~Rev.~Lett.~103, 210401 (2009)]. 
	We reveal that the nature of projective measurements in quantum mechanics leads to a fundamental, nontrivial bias in this measure. 
	We observe and study the functional dependence of this bias to permit a demonstration of applications of local quantum probing. 
	An extension of our approach can act as a versatile reference, relevant for understanding complex systems.
\end{abstract}


\date{\today}

\maketitle

In nature any quantum system inevitably interacts with its 
environment\,\cite{breuer_theory_2002}. 
This interaction induces dynamics, which creates classical and quantum correlations, and will eventually lead to decoherence and dissipation for observables of the \emph{open} system. 
Common approaches to enable a description of the open system dynamics involve the approximation of a Markovian process, i.e., a memoryless time evolution. 
However, in many cases this assumption is not justified, and distinct dynamical features witness underlying non-Markovian behavior.

The classical definition of non-Markovianity fails in the quantum regime, due to the special role of measurements, as described by the projection postulate\,\cite{breuer_textitcolloquium_2016-1}.
Recently, several definitions of quantum non-Markovianity, as 
well as quantitative measures, have been developed\,
\cite{wolf_assessing_2008,breuer_measure_2009-2,rivas_entanglement_2010,chruscinski_measures_2011-6,luo_quantifying_2012-1,chruscinski_degree_2014-4}, see reviews\,\cite{rivas_quantum_2014,breuer_textitcolloquium_2016-1,de_vega_dynamics_2017}.
The physical implications of memory effects have initiated a variety of applications for 
diverse quantum systems and phenomena, e.g., Ising/Heisenberg spin chains and Bose-Einstein 
condensates\,\cite{apollaro_memory-keeping_2011,haikka_quantifying_2011,haikka_non-markovianity_2012}, quantum phase transitions\,\cite{gessner_observing_2014}, Anderson localization\,\cite{lorenzo_quantum_2016}, opto-mechanical systems\,\cite{groblacher_observation_2015-1}, chaotic systems\,\cite{znidaric_non-markovian_2011,pineda_measuring_2016}, quantum dots\,\cite{madsen_observation_2011}, energy transfer processes in photosynthetic complexes\,\cite{rebentrost_communication:_2011}, and quantum metrology\,\cite{chin_quantum_2012}. 

The definition of quantum non-Markovianity developed in Ref.\,\cite{breuer_measure_2009-2} features a physical interpretation based on concepts of quantum information theory. It employs a distance measure in state space to characterize the distinguishability of quantum states\,\cite{nielsen_quantum_2000} of the open system. 
In this context, non-Markovianity is identified as backflow of information to the system, i.e., as an increase in distinguishability.
So far, non-Markovianity and related questions of initial system-environment correlations have been experimentally observed in photonic\,\cite{liu_experimental_2011-1,li_experimentally_2011,smirne_experimental_2011,cialdi_two-step_2014,bernardes_experimental_2015,tang_experimental_2015,sun_non-markovian_2016-1}, nuclear magnetic resonance\,\cite{bernardes_high_2016-1}, and trapped-ion systems\,\cite{gessner_local_2014-2}.

Trapped atomic ions are well suited to further investigate aspects of memory effects.
Individual control of electronic and motional degrees of freedom permit the realization of effective spins with tunable couplings via or to bosonic degrees of freedom\,\cite{wineland_nobel_2013,leibfried_quantum_2003,blatt_entangled_2008,porras_mesoscopic_2008,schneider_experimental_2012,clos_time-resolved_2016}.
Techniques for preparation, (coherent) manipulation, effective interaction, and detection of quantum states are performed with efficiencies close to unity\,\cite{leibfried_quantum_2003,blatt_entangled_2008,ballance_hybrid_2015,kienzler_quantum_2015,harty_high-fidelity_2014}.
Isolation from surroundings approximates a closed system with parameters that can be tuned continuously---from a simple toylike system, still allowing for exact numerical treatment of pure and mixed states, up to complex system-environment configurations and interactions\,\cite{barreiro_open-system_2011,porras_mesoscopic_2008,schneider_experimental_2012,monroe_scaling_2013,ramm_energy_2014,mielenz_arrays_2016}.  

In this work, we study fundamental aspects of quantum non-Markovianity in a trapped-ion system. 
We experimentally monitor and quantify the exchange of information between an open system and its well-defined quantum environment with the measure defined in Ref.\,\cite{breuer_measure_2009-2}.
Thereby, we reveal that its mathematical definition translates intrinsic uncertainties into a systematic bias.
Still, we show how to employ our system as a local quantum probe to characterize system-environment couplings and environmental states.

To define quantum non-Markovianity for a system $S$ interacting with its environment $E$, the authors of Ref.\,\cite{breuer_measure_2009-2} suggest to utilize the time evolution of the trace distance, defined by $D(t)=\frac{1}{2}||\rho_S^1(t)-\rho_S^2(t)||$. It quantifies the distinguishability of two system states $\rho_S^{1,2}$\,\cite{nielsen_quantum_2000}, which are obtained by tracing out the environmental degrees of freedom.
While Markovian processes are defined by a monotonic decrease of $D(t)$, the 
characteristic feature of non-Markovian dynamics is any
increase of $D(t)$\,\cite{breuer_measure_2009-2}.
Further, the accumulated growth of $D$ within a maximal duration $t_\text{max}$, where $D$ is 
sampled in steps of $\Delta t$, is quantified by\,\cite{breuer_measure_2009-2}:
\begin{equation} 
\label{NM-measure}
\mathcal{N} = \sum_{t=\Delta t}^{t_\text{max}} \;\left[ D(t) - D(t-\Delta t) \right]_{>0}.
\end{equation}
Explicitly, the sum extends over all positive changes of $D(t)$. In the following, we consider the non-Markovianity corresponding to a representative pair of orthogonal initial states ${\rho_S^{1,2}(t=0)}$.
We note that the choice of the sampling rate $\gamma \equiv 1/\Delta t$ and $1/t_\text{max}$ defines the highest and lowest frequency, respectively, with which a growth in $D$ can be detected.

In classical probability theory, there exists a mathematical condition for stochastic processes to be Markovian in terms of conditional probability distributions\,\cite{van_kampen_stochastic_1992}. 
This definition cannot be transferred to the quantum regime, as the quantum state changes discontinuously and randomly conditioned on the outcome of a projective measurement. In particular, measurements on the open system completely destroy all---classical and quantum---correlations between system and environment. Hence, they strongly influence the subsequent dynamics\,\cite{breuer_textitcolloquium_2016-1}. 
On the one hand, Eq.\,\eqref{NM-measure} provides a clear definition for a measure of quantum non-Markovianity which is independent of measurement-induced state changes described by the projection postulate.
On the other hand, measurements are subjected to intrinsic uncertainties, referred to as \emph{quantum projection noise} (QPN)\,\cite{itano_quantum_1993}: Consider a superposition state of a two-level system, $\ket{\psi}\equiv c_A\ket{A}+c_B\ket{B}$, with $|c_A|^2+|c_B|^2 =1$. Any projective measurement transfers $\ket{\psi}$ into the pointer basis of the measurement device. For example, if the pointer basis is $\{\ket{A},\ket{B}\}$, the result indicates either $\ket{A}$ or $\ket{B}$, with probability $|c_A|^2$ or $|c_B|^2$, respectively. Consequently, expectation values can only be determined by averaging $r$ repetitions. The related statistical uncertainty is proportional to $1/\sqrt{r}$ and persists, even in absence of any uncertainty in state preparation.
We point out that the mathematical definition of $\mathcal{N}$ translates QPN into a systematic bias $\mathcal{B}$.
This yields the explicit functional dependence $\mathcal{N} = \mathcal{N}(t_\text{max},\gamma,r)$. We regard values with zero QPN and infinite $\gamma$ as \emph{true} values, i.e., $\mathcal{N}_\text{true}\equiv \lim_{\gamma,r\to\infty}\mathcal{N}(t_\text{max},\gamma,r)$.
We identify $\mathcal{B} \equiv \mathcal{N} - \mathcal{N}_\text{true}$ to be a nontrivial function of the particular evolution $D(t)$ and the parameters $t_\text{max}$, $\gamma$, and~$r$.

\begin{figure}
	\includegraphics{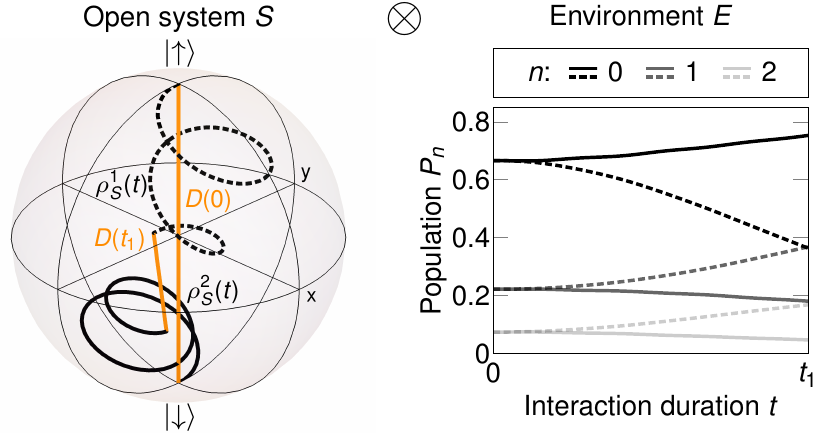}
	\caption{
		Toy system to study quantum memory effects.
		Illustration of the total system, composed of a single spin-1/2, initially in $\rho_S^1(0) = \ket{\uparrow}\bra{\uparrow}$ (dashed) or $\rho_S^2(0) = \ket{\downarrow}\bra{\downarrow}$ (solid), and a bosonic environment, initially in a thermal state with low $\bar{n}$. 
		While $\rho_S^{1,2}(t)$ are depicted on the left in the Bloch-sphere representation, the related populations of the $n=\{0,1,2\}$ environmental states are shown on the right. 
		Information is transferred from $S$ to $E$ and into correlations/entanglement (not depicted);
		the amount is accounted for by the change of distinguishability~$D(t)$ of $\rho_S^{1,2}(t)$.
		\label{fig1}}
\end{figure}

In order to investigate properties of quantum non-Markovian dynamics, we consider the following toy system.
It is composed of a single spin-1/2, representing the open system $S$, and a bosonic degree of freedom that spans its environment $E$, see Fig.\,\ref{fig1}. 
The bipartite system $S+E$ is assumed to be isolated from an additional surrounding $X$.
We write the open system's Hamiltonian as $H_S = \hbar \omega_z\sigma_z/2$, where $\sigma_z$ is the Pauli matrix with eigenstates $\ket{\downarrow}$ and $\ket{\uparrow}$ and effective energy splitting $\hbar \omega_z$, and the reduced Planck constant $\hbar$. 
The environment is represented by the Hamiltonian $H_E = \hbar \omega_E a^\dagger a$, with annihilation (creation) operators $a$ ($a^\dagger$) and eigenfrequency $\omega_E$, and the Fock states are labeled $n$. 
The dynamics of the total system $S + E$ is governed by the Hamiltonian\,\cite{porras_mesoscopic_2008}
\begin{eqnarray}
H &=& H_S + H_E + H_I \\
&=&
\frac{\hbar\omega_z}{2} \sigma_z +
\hbar\omega_E a^\dagger a 
+ \frac{\hbar\Omega}{2}\left[
\sigma^+{\text{e}}^{i \eta ({a}^\dagger + a)} +h.c.
\right].\nonumber
\label{fullham}
\end{eqnarray}
Here, we express the interaction term $H_I$ by the spin coupling rate $\Omega$, spin-flip operators $\sigma^\pm \equiv (\sigma_x \pm i \sigma_y)/2$, Pauli matrices $\sigma_{x,y}$, and the spin-boson coupling-parameter $\eta$. 
Further, we investigate the evolution of initial product states $\rho (0) = \rho_S(0) \otimes \rho_E(0)$, with two representative states, $\rho_S^1(0) \equiv \ket{\uparrow}\bra{\uparrow}$ and $\rho_S^2(0) \equiv \ket{\downarrow}\bra{\downarrow}$, and thermal states $\rho_E(0)$ defined by average occupation numbers $\bar{n}$.
We choose $\rho_E(0)$ near the ground state to ensure that energies of spin and bosonic degrees of freedom remain comparable, enabling observations of distinct features of quantum memory. In Figure\,\ref{fig1}, we illustrate an exemplary time evolution $\rho_S^{1,2}(t)$ and changes in Fock-state populations that indicate a transfer of information from $S$ to $E$.

In our experiment, we implement $H$ with a single trapped $^{25}$Mg$^+$. For all our measurements, we ensure that residual decoherence rates $\Gamma_\text{dec}$, due to couplings to $X$ (technical noise), are negligible, $\Gamma_\text{dec} \ll 1/t_\text{max} < \Omega$\,\cite{supplement}.
Two electronic hyperfine states form $S$, while $E$ is composed of a motional mode with frequency $\omega_E/(2\pi) = 1.920(3)\,\text{MHz}$. 
The coherent $S$-$E$ interaction $H_I$ is implemented via two-photon stimulated Raman transitions\,\cite{leibfried_quantum_2003} with $\Omega/(2\pi) \approx 100\,\text{kHz}$ and $\eta\approx0.32$. 
More details on the experimental implementation and data analysis are described in the Supplemental Material\,\cite{supplement} and Refs.\,\cite{schneider_experimental_2012,clos_time-resolved_2016}. 
To record $D(t)$, we perform measurement series of time-resolved spin-state tomography\,\cite{leibfried_quantum_2003}. 
Each sequence starts with initialization of $\rho_S^1(0)$ or $\rho_S^2(0)$, with dedicated $\bar{n}$.
We implement $H$ for variable duration $t \in \left[0,9\tau\right]$, with $\tau\equiv2\pi/\Omega$.
Subsequently, we detect expectation values $\langle\sigma_l(t)\rangle$, $(l=x,y,z)$, in individual sequences for each $l$ with fixed $r=r_0\equiv500$ and $\gamma=\gamma_0 \approx 15 \tau^{-1}$. 
From recorded $\langle \sigma_l(t)\rangle$, we determine $\rho_S^{1,2}(t)$, corresponding $D(t)$ and $\mathcal{N}$, and their statistical uncertainties\,\cite{supplement}.
To assess systematic effects of our measurements, we compare our data with numerical simulations of the total system dynamics generated by $H$\,\cite{supplement, clos_time-resolved_2016}. We conduct independent calibration measurements to determine corresponding parameters $\omega_E$, $\omega_z$, $\Omega$, and $\bar{n}$.
In particular, we choose $\gamma$ and $r$ according to our experimental realizations to generate numerically simulated values for the averages $\langle \sigma_l\rangle$. These yield the dispersion of $D$ and values $\mathcal{N}$ that include the effect of QPN\,\cite{supplement}. 
Additionally, to estimate $\mathcal{N}_\text{true}$ and, therefore, to quantify $\mathcal{B}$, we perform numerical simulations. To this end, we consider zero noise amplitude, equivalent to $r \rightarrow \infty$, a sampling rate $100\, \gamma_0$, and all other parameters fixed according to the experimental realizations\,\cite{supplement}.

\begin{figure}
	\includegraphics{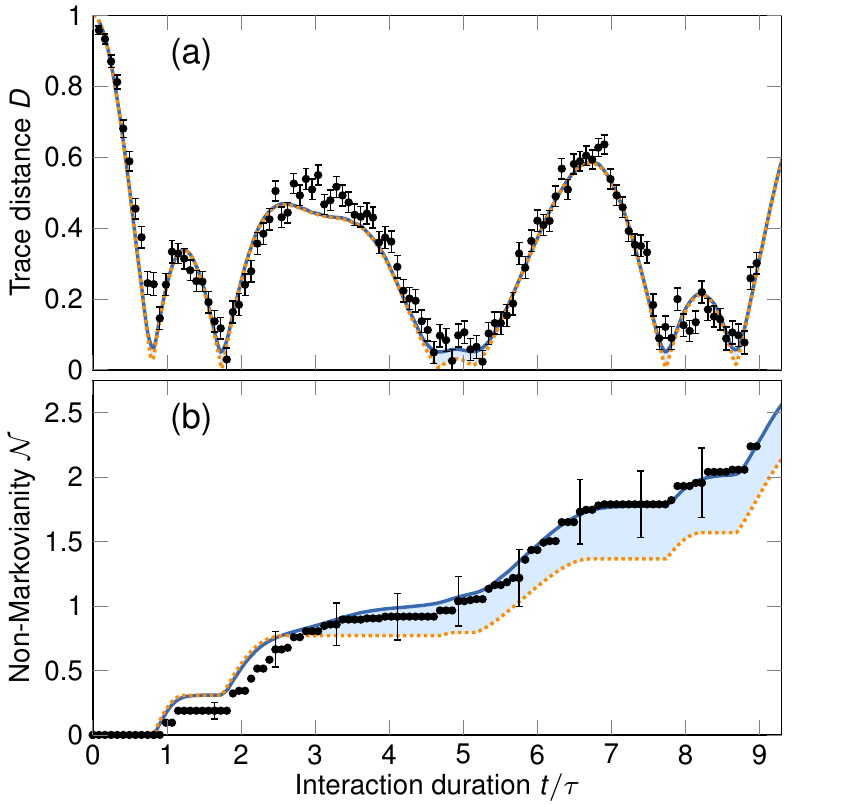}
	\caption{
		Features of non-Markovianity.
		Experimental results (data points) for $D(t)$ and $\mathcal{N}$, for $\bar{n} = 1.0(1)$ and $\omega_z \approx \omega_E$, are compared to corresponding (solid lines) and true (dotted lines) numerical results. 
		Shaded areas indicate the effect of the fundamental quantum projection noise (QPN).
		(a)\,Non-Markovian behavior is indicated by increases of $D$. 
		Systematic deviations of data points (error bars: 1\,s.d.) from the solid line reveal technical imperfections, while systematic deviations due to QPN remain negligible.
		(b)\,Memory effects are evidenced by increasing $\mathcal{N}$. Noise yields an increasing bias, predominantly, when amplitudes in the dynamics become comparable to QPN amplitudes. Error bars (1\,s.d.) depict correlated statistical uncertainties, and we show representatives only\,\cite{supplement}.
		\label{fig2}}
\end{figure}

First, we consider an example to discuss distinct features of recorded non-Markovian behavior.
In Figure\,\ref{fig2}, we show measured $D$ and $\mathcal{N}$ (data points) for $\bar{n}=1.0(1)$ and resonant interaction, $\omega_z/\omega_E=1.000(2)$, and find good agreement with numerical simulations (solid lines).
Error bars depict the amount of QPN, while additional experimental uncertainties are neglected.
Information, initially encoded in $S$, is transferred to $E$ or $S$-$E$ correlations, evidenced by decreasing $D$ and flat $\mathcal{N}$. 
Memory effects are witnessed whenever $D$ increases, accounted for by an increase of $\mathcal{N}$.
The estimated true numerical results (dotted lines) deviate from data in $D$ only for particular durations, indicating residual systematic/technical effects. 
In contrast, they increasingly deviate from data in $\mathcal{N}$.
We find that QPN accumulates to a systematic bias in $\mathcal{N}$. Predominantly, the increases of $\mathcal{B}$ occur for durations of near constant $\mathcal{N}$. Here, amplitudes of the dynamical evolution of $D$ become comparable to noise amplitudes in $D$ that are a direct consequence of QPN. 

\begin{figure}
	\includegraphics{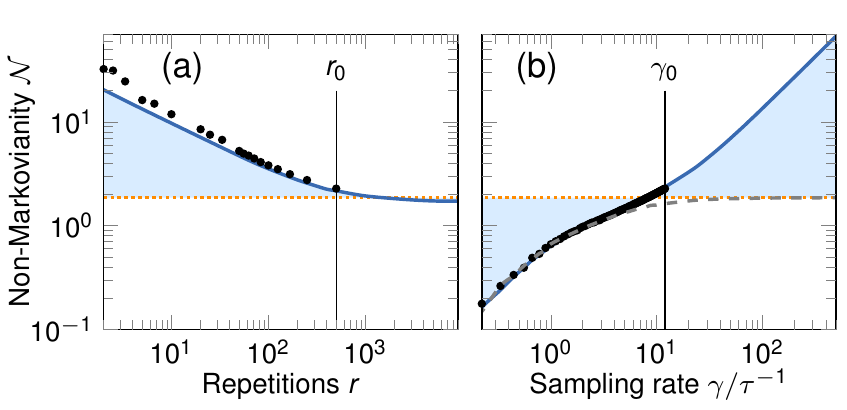}
	\caption{
		Characterizing the impact of QPN and sampling.
		Experimental results $\mathcal{N}(\gamma,r)$ (data points) are compared to corresponding (solid lines) and true (dotted lines) numerical results for $t_\text{max}=9 \tau$ and $D$ shown in Fig.\,\ref{fig2}(a).
		The bias $\mathcal{B}(\gamma,r)$ is indicated by the shaded area, while error bars are omitted for clarity.
		(a)\,We find that $\mathcal{B}(\gamma_0, r_0)/\mathcal{N}_\text{true} \approx + 17\%$ and substantially increases for $r<r_0$ due to QPN. 
		The bias approaches negative values for $r>r_0$, e.g., $\mathcal{B}(\gamma_0, r=10^4)/\mathcal{N}_\text{true} \approx - 8\%$, indicating that low amplitudes of fast dynamics in $D$ are missed by our recording at $\gamma_0$.
		(b)\,QPN leads to $\lim_{\gamma\rightarrow\infty}\mathcal{B}(\gamma,r_0)\rightarrow\infty$, while finite sampling yields negative $\mathcal{B}$ for $\gamma\tau<8$.
		The dashed line illustrates $\lim_{r\rightarrow\infty}\mathcal{N}(\gamma,r)$. 
		\label{fig3}}
\end{figure}

Next, we present results to investigate $\mathcal{N}(\gamma,r)$ for fixed $t_\text{max} = 9 \tau$.
In Figure\,\ref{fig3}, we quantify the impact of QPN on $\mathcal{N}$ as a function of $r$ and $\gamma$, for the time evolution depicted in Fig.\,\ref{fig2}.
\begin{figure*}
	\includegraphics{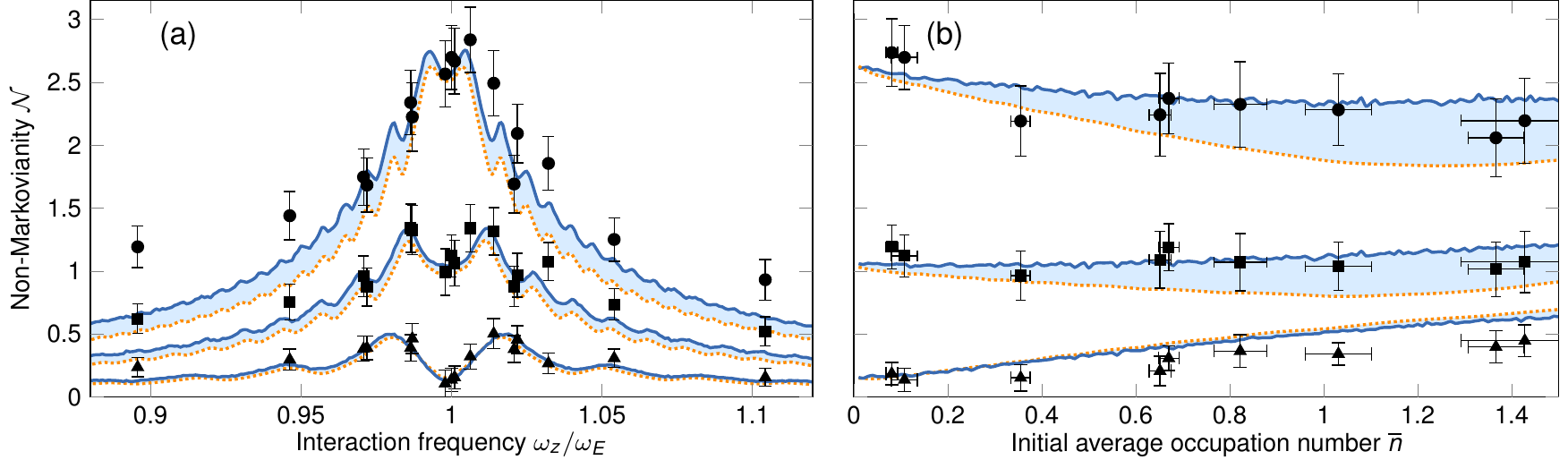}
	\caption{
		Application of local quantum probing.
		Experimental results (data points, error bars: 1\,s.d.) of $\mathcal{N}(\gamma_0,r_0)$ for distinct $t_\text{max}=\{2,5,9\}\tau$ (triangles, squares, circles) are compared to corresponding (solid lines) and true (dotted lines) numerical results.
		(a)\,We choose $\bar{n}=0.09(2)$, to ensure elementary quantum dynamics. In turn, corresponding $D(t)$ are close to trivial, i.e., sinusoidal with variable frequency and amplitude.
		Contributions of QPN (shaded areas) are near constant.
		(b)\,We observe positive and negative slopes in $\mathcal{N}(\bar{n})$ for $\omega_z/\omega_E=1.000(2)$, reflecting increasingly complex $S$-$E$ dynamics.
		Here, the impact of QPN is variable.
		\label{fig4}}
\end{figure*}
We vary $r$ for fixed $\gamma=\gamma_0$ by postselection of random subensembles of the $r_0$ experimental realizations, generate resampled evolutions $D(t)$, and calculate corresponding $\mathcal{N}(\gamma_0,r)$\,\cite{supplement}. 
The results, depicted in Fig.\,\ref{fig3}(a), are in agreement with corresponding simulations.
For increasing $r$, the results approach the estimated $\mathcal{N}_\text{true}$. However, for $r\gg r_0$, we find a significant underestimation, e.g., $\mathcal{B}(\gamma_0, r=10^4)/\mathcal{N}_\text{true} \approx - 8\%$. We attribute this to dynamical features that cannot be resolved due to our choice of $\gamma_0$.
Further, we vary the mean sampling rate by random postselection of data points in recorded $D$ for fixed $r=r_0$\,\cite{supplement}, and show corresponding results and numerical simulations in Fig.\,\ref{fig3}(b). 
We find that $\mathcal{N}(\gamma,r_0)$ continues to increase with increasing $\gamma$, and even diverges for $\gamma \gg \gamma_0$. 
Sampling rates that are too small to resolve fast dynamical features in $D$ result in a significant underestimation of $\mathcal{N}$, i.e., $\mathcal{B}<0$. However, sampling rates that approach reasonable values overestimate $\mathcal{N}_\text{true}$ due to the contribution of QPN and, correspondingly, $\lim_{\gamma \rightarrow \infty}\mathcal{B}(\gamma,r_0)\rightarrow\infty$. 
Generally, the variable amount of $\mathcal{B}$ can hamper a determination of~$\mathcal{N}$.

Nevertheless, we demonstrate that an application of the non-Markovianity measure $\mathcal{N}$ to probe characteristics of $E$, cf. Ref.\,\cite{breuer_textitcolloquium_2016-1}, is still possible, when QPN is taken into account.
In particular, we keep our parameters of $S$ and $E$ close to the simple regime and benchmark the ability of $S$ to probe $S$-$E$ couplings and variable states of $E$.
In a first measurement series, we probe changes in $S$-$E$ couplings by variation of $\omega_z$ near $\omega_E $, for fixed $\bar{n}=0.09(2)$, and determine $\mathcal{N}(\gamma_0,r_0)$. 
Figure\,\ref{fig4}(a) depicts $\mathcal{N}(\gamma_0,r_0, \omega_z)$ for three distinct $t_\text{max}=\{2,5,9\}\tau$.
We observe resonances near $\omega_z \approx \omega_E$ that differ significantly in shape, depending on $t_\text{max}$. 
For small $t_\text{max}$, recorded  $\mathcal{N}$ feature a double-peak structure. 
This reflects an expected increase of the effective coupling rate $\Omega' $ (faster dynamics) for a detuning from resonance by $\delta\omega_z \equiv \omega_z-\omega_E$, which can be estimated by $\Omega' \propto \sqrt{\Omega^2 + \delta\omega_z^2}$.
In contrast, for larger $t_\text{max}$, line shapes become dominated by the resonant $S$-$E$ interaction, since amplitudes in $D(t)$ are $\propto \Omega^2/\Omega'^2$, i.e., they are largest for $\omega_z=\omega_E$, cf.\,\cite{supplement}.
Comparing our data to corresponding and true numerical results, we estimate $\mathcal{B}/\mathcal{N}_\text{true} \approx 18\%$ on average with small variations, and we can experimentally resolve predicted features.
In a second series of measurements, we probe the environmental state by tuning the initial $\bar{n}$.
In Figure\,\ref{fig4}(b), we depict experimental and numerical results of $\mathcal{N}(\gamma_0,r_0, \bar{n})$ for $t_\text{max}=\{2,5,9\}\tau$ and fixed $\omega_z \approx \omega_E$, and compare them to the related true values.
For short durations, we find an increase of $\mathcal{N}(\bar{n})$ for increasing $\bar{n}$, while for longer durations this trend changes and true values suggest a decrease of $\mathcal{N}(\bar{n})$. 
We reveal that $|\mathcal{B}/\mathcal{N}_\text{true}|$ varies substantially between 0\% and 45\%, depending on $t_\text{max}$ and $\bar{n}$.
For increasing $\bar{n}$, the spin interacts with a larger number of Fock states and the evolution of $D$ features a less trivial frequency spectrum\,\cite{supplement}. 
This results in a faster and more complex dynamics $D(t)$ that cannot be resolved with constant significance for fixed $\gamma_0$ and $r_0$.

While our results can be considered to be a textbook example, for increasingly complex or unknown states of $E$ and/or interactions of $S$ and $E$, the quantification of $\mathcal{B}$ becomes less trivial. Leaving the regime of numerical tractability, it can be successful to experimentally record functions $\mathcal{N}(\gamma,r)$, as presented above. An estimation for $\mathcal{B}(\gamma_0,r_0)$ may be obtained by extrapolating $\lim_{r\rightarrow\infty}\mathcal{N}(\gamma_0,r)$ for varying $\gamma$.
A detailed mathematical characterization of $\mathcal{B}(\gamma,r)$ can assist such extrapolations.
In addition, an estimation of $\mathcal{B}$ can be achieved, e.g., by optimizing semi-empirical models, which are designed to describe $D$ and tested to return recorded $\mathcal{N}(\gamma,r)$\,\cite{supplement}. 
Further, in combination with these models, Fourier series of recorded $D(t)$ may allow to isolate dynamics of interest from QPN. Thereby, frequency filters or regularization methods may allow to estimate $\mathcal{B}(\gamma,r)$\,\cite{press_numerical_2007}.
But, we point out that an optimized strategy depends on technical limitations, excess noise levels, and properties of $S$ and $E$.

In summary, we set up our trapped-ion system to implement an effective spin representing an open system, which we couple to an environment composed of a bosonic degree of freedom. 
We investigate the evolution of the trace distance of two initially orthogonal spin states to study distinct features of quantum memory effects. 
Our results demonstrate that inherent fluctuations, arising from random projection during the measurement process, yield not only uncertainties, but also a significant bias in the quantification of such effects. This affects any experimental platform and even numerical approaches, such as Monte Carlo simulations.
We quantify this bias in our system to determine accurate values of the quantum non-Markovianity measure. 
On the basis of these findings, we employ the open system as a local quantum probe to explore characteristics of system-environment couplings and environments.
Our experimental platform is ideal to tune to more complex environments and couplings\,\cite{clos_time-resolved_2016}, which includes adding spin or bosonic degrees of freedom, preparing a variety of initial environmental states, and engineering couplings to additional, even classical, surroundings.
As such, it can act as a versatile reference and aids understanding of physical systems, in which parameters are less controlled, and other noise sources contribute substantially to an excess bias.

Further, our fundamental findings lead to questions concerning generalizations and applications of non-Markovianity. 
The effect of QPN on other measures, which are based on, e.g., the divisibility of the dynamical map\,\cite{rivas_entanglement_2010,chruscinski_measures_2011-6,chruscinski_degree_2014-4} or the mutual information between the open system and an ancilla system\,\cite{luo_quantifying_2012-1}, needs to be studied, as we expect them to be significantly influenced by QPN as well.
Overall, it may be practical to extend definitions of non-Markovianity measures by including physical constraints, in order to enable a comparison of different systems. 
Time scales and the related flow of exploitable information may depend on the application.
In any case, an upper limit for the sampling rate may be given, e.g., by the so-called quantum speed limit\,\cite{deffner_quantum_2013}, which, in turn, would directly limit the impact of QPN.
Based on such extensions, applications for characterizing time scales and experimentally accessible complexity measures may emerge.
These are needed, e.g., in the context of equilibration dynamics and thermalization in isolated quantum systems\,\cite{clos_time-resolved_2016}.

\begin{acknowledgments}
	We thank D. Porras for providing the software package used to perform numerical simulations and T. Filk for helpful comments on the manuscript. Our work was supported by the Deutsche Forschungsgemeinschaft [SCHA 973; 91b (INST 39/828-1 and 39/901-1 FUGG)] as well as the European Union (EU) through the Collaborative Project QuProCS (Grant Agreement No. 641277).
\end{acknowledgments}

\clearpage
{\section{\large Supplemental Material}}

\section{Experiments}

We employ a linear radio-frequency (rf) Paul trap with drive frequency $\Omega_\text{rf}/(2\pi) \approx 56\,\text{MHz}$ to trap single ${}^{25}\text{Mg}^+$ with secular frequencies $\omega_\text{rad}/(2\pi)\approx \{3.9,4.7\}\,\text{MHz}$ in radial directions and
$\omega_E/(2\pi)\approx 1.9\,\text{MHz}$ in axial direction. We implement control of the internal and external states of the ion with coherent laser transitions. Details on the experimental setup can be found in the Supplemental Material of Ref.\,\cite{clos_time-resolved_2016} and in Refs.\,\cite{clos_decoherence-assisted_2014,schaetz_towards_2007}.
Two electronic ground states in the hyperfine manifold of $^{25}$Mg$^+$ (nuclear spin 5/2) form our open system, $|{\uparrow}\rangle \equiv 3S_{1/2}|F{=}3,m_F{=}3\rangle$ and $|{\downarrow}\rangle \equiv 3S_{1/2}|F{=}2,m_F{=}2\rangle$, with a splitting of about $2\pi\, 1775\,\text{MHz}$, where $F$ and $m_F$ denote the total angular momentum quantum numbers of the ion's valence electron. 

\begin{figure}[htb]
	\centering
	\includegraphics[width=8.6cm]{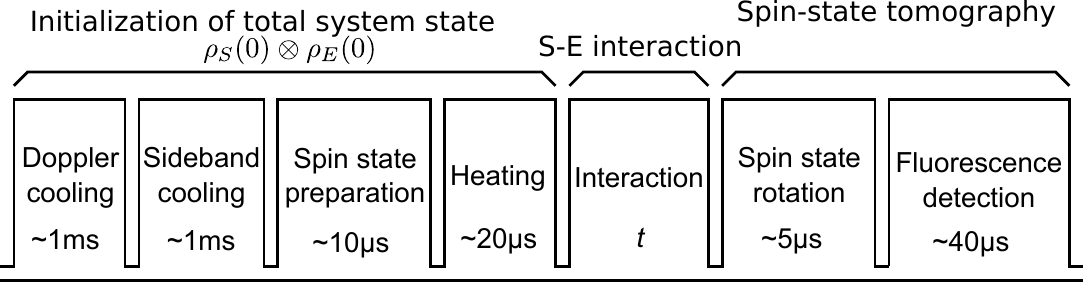}
	\caption{
		Experimental sequence (not to scale) implemented to record dynamics of $\langle\sigma_l(t)\rangle$, for $l=x,y,z$. 
		In particular, we adjust relative phases of the laser fields, which also implemente the $S$-$E$ interaction, within the analysis pulse to enable projection of the different spin components onto the $z$ axis. We perform individual sequences to observe $\langle\sigma_l(t)\rangle$ with interaction durations $t \in \left[0,9\tau\right]$ varied in steps of $\Delta t \approx \tau/15$.
		\label{figsequence}
	}
\end{figure}

To analyze the open system's dynamics under a given interaction with its quantum environment, we prepare factorizing states of the total system, $\rho_S(0) \otimes \rho_E(0)$, turn on the interaction for variable duration $t$, and, subsequently, perform spin-state tomography.
A schematic of the experimental sequence is depicted in Fig.\,\ref{figsequence}. 

We employ Doppler cooling and subsequent resolved sideband cooling to prepare the ion in a thermal motional state with $\bar{n} \approx 0.1$. 
The ion's spin state $\rho_S$ is initialized via optical pumping and microwave electron shelving techniques\,\cite{leibfried_quantum_2003} to prepare either $\rho_S^1(0) = \ket{\uparrow}\bra{\uparrow}$ or $\rho_S^2(0)= \ket{\downarrow}\bra{\downarrow}$.
To initialize motional states $\rho_E(0)$ of interest with $\bar{n} \in \left[0,1.5\right]$, we apply electric-field white-noise to the ion via one of our trap electrodes.
This results in a controlled heating rate $\dot{\bar{n}} \approx 0.15\,\text{quanta}/\mu\text{s}$, while relevant ambient heating rates are about $6\,\text{quanta/s}$.

Following state initialization, we turn on the system-environment interaction for variable duration $t$ via two-photon stimulated Raman (TPSR) transitions\,\cite{leibfried_quantum_2003} with variable two-photon detuning $\omega_z$\,\cite{porras_mesoscopic_2008}. 
The laser beams are detuned from the transition to $3P_{3/2}$ by $\Delta_R/(2\pi) \approx 65\,\text{GHz}$ and aligned such that the effective wave vector points along the axial direction. Intensities of the beams are adjusted to yield the coupling strength $\Omega/(2\pi) \approx 100\,\text{kHz}$. We choose interaction durations $t \in \left[0,9\tau\right]$ in steps of $\Delta t \approx \tau/15$.

\begin{figure}[htb]
	\centering
	\includegraphics[]{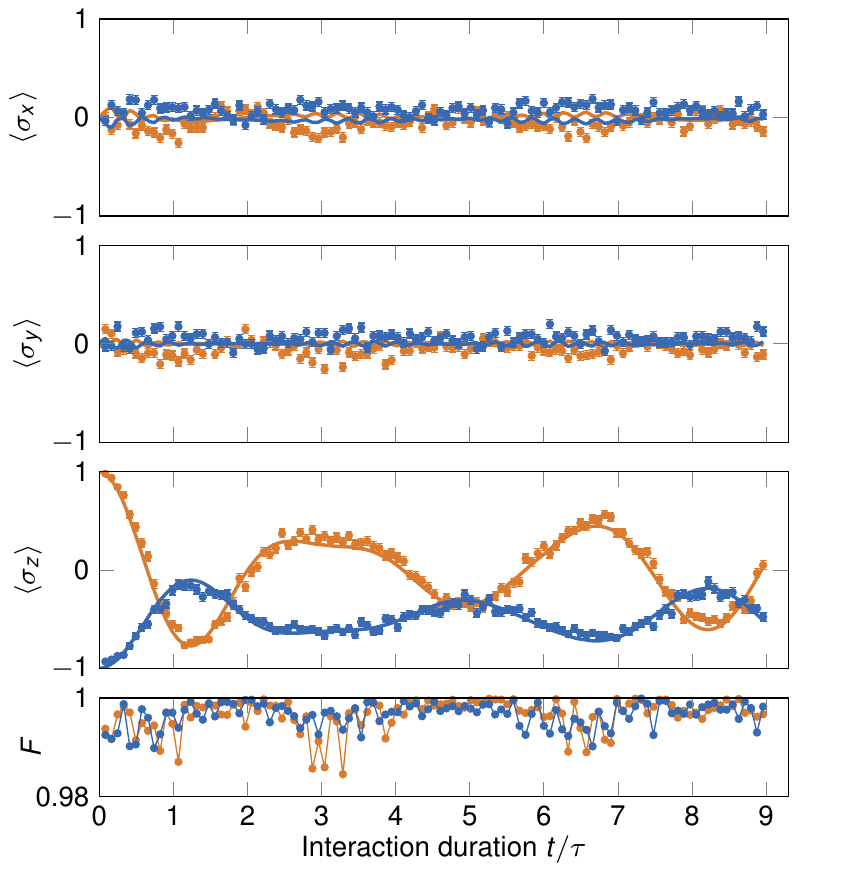}
	\caption{
		Measurement of $\rho_S^{1,2}(t)$ and corresponding fidelity $F(t)$. We compare measurements of spin-state components $\langle \sigma_l(t) \rangle\;(l=x,y,z)$ (data points) with simulated $\langle \sigma_l(t) \rangle_\text{true}$ (solid lines) for both initial states $\rho_S^1$ (orange) and $\rho_S^2$ (blue) for $D(t)$ from Fig.\,2.
		We observe small systematic variations of our experimental fidelity that originate from the motional-state sensitivity of our tomography scheme.
		\label{figtomography}
	}
\end{figure}

To determine $D(t)$, we measure $\langle \sigma_l(t) \rangle\;(l=x,y,z)$ for both initial spin states $\rho_S^{1,2}$, see Fig.\,\ref{figtomography}.
We implement state-sensitive fluorescence detection\,\cite{leibfried_quantum_2003} via the closed cycling transition $\ket{\uparrow}\leftrightarrow3P_{3/2}\ket{4,4}$ in order to obtain occupation probabilities $P_{\ket{\uparrow}} \in \left[0,1\right]$. This corresponds to expectation values $\langle \sigma_z \rangle = 2P_{\ket{\uparrow}} - 1 \in \left[-1,+1\right]$. To detect $\langle\sigma_{x,y}\rangle$, we map the Bloch-vector component of interest onto the $z$ axis via well-defined rotation pulses employing TPSR transitions with $\omega_z=0$, such that $P_{\ket{\uparrow}}$ yields expectation values $\langle \sigma_{x,y} \rangle$.
Each sequence is repeated $r= r_0 \equiv 500$ times to yield averages $\langle \sigma_l(t) \rangle$. Corresponding uncertainties $\delta \langle \sigma_l(t)\rangle$\,\cite{itano_quantum_1993}, representing (only) quantum projection noise (QPN), are given by the standard deviation of the binomial distribution:
\begin{equation}
	\delta \langle \sigma_l \rangle = 2\, \sqrt{\frac{1}{r}\left(\frac{\langle\sigma_l\rangle+1}{2}\right)\left(1-\frac{\langle\sigma_l\rangle+1}{2}\right)}.
	\label{qpnuncertainty}
\end{equation}   
From the measured $\langle\sigma_l(t)\rangle$, we calculate Bloch vectors $\mathbf{v}(t)=\left(\langle \sigma_x(t) \rangle,\langle \sigma_y(t) \rangle,\langle \sigma_z(t) \rangle\right)^T$, and the corresponding trace distance of the states $\rho_S^{1,2}$
\begin{equation}
	D(\mathbf{v_S^1},\mathbf{v_S^2}) = \frac{1}{2} \sqrt{\left(\mathbf{v_S^1}-\mathbf{v_S^2}\right)^\dagger \cdot \left(\mathbf{v_S^1}-\mathbf{v_S^2}\right)}\nonumber
\end{equation}
with statistical uncertainties, obtained by error propagation:
\begin{equation}
	\delta D = \sqrt{\sum_{l=x,y,z}\sum_{m=1,2\vphantom{yl}} \left(\frac{\partial D}{\partial \langle \sigma_l^m\rangle} \delta \langle \sigma_l^m \rangle \right)^2}.
\end{equation}
We get $\mathcal{N}$ by evaluating changes between two consecutive measurements, $\Delta D(t) \equiv D(t) - D(t-\Delta t)$, and calculating 
\begin{equation}
	\mathcal{N} = \sum_{t=\Delta t}^{t_\text{max}} \left.\Delta D(t)\right|_{>0}. \nonumber
\end{equation}
The sum extends over all positive values of $\Delta D(t)$, i.e., all increases $\Delta D_i \equiv D_i - D_{i-1} > 0$. The corresponding uncertainty $\delta \mathcal{N}$ is again obtained by error propagation of the contributing uncertainties $\delta \Delta D_i = \sqrt{\delta D_i^2 + \delta D_{i-1}^2}$ and consequently:
\begin{equation}
	\delta \mathcal{N} = \sqrt{ \sum_{i} \delta\Delta D_{i}^2}. \nonumber
\end{equation}
The above expression considers only statistical uncertainties, while systematic effects are not considered. However, we note that our tomography scheme is based on well-defined spin-state rotations, that are most efficiently performed for motional states with $\bar{n} \rightarrow 0$. Since Rabi rates depend on $n$\,\cite{wineland_experimental_1998}, the fidelity of our experimental sequence is a function of $\rho_E$. 
We check the fidelity $F$ of our experimental sequence by evaluating
\begin{equation}
	F \left[\rho_\text{exp}(t),\rho_\text{num}(t)\right] = \sqrt{ \sqrt{\rho_\text{exp}(t)}\rho_\text{num}(t) \sqrt{\rho_\text{exp\vphantom{p}}(t)}},\nonumber
\end{equation}
where $\rho_\text{exp}$ represents the experimentally realized/detected spin state and $\rho_\text{num}$ is numerically simulated (see next section) for the corresponding experimental parameters. Figure\,\ref{figtomography} depicts measured $\langle\sigma_l(t)\rangle$ compared to numerical simulations and the corresponding fidelity $F(t)$ for the time evolution depicted in Fig.\,2. 

We observe small systematic modulations of $F(t)$ as $\rho_E$ is modified during the interaction, see Fig.\,\ref{figtomography}. 
For the measurement of $D(t)$ with $\bar{n} = 1.0(1)$, see Fig.\,\ref{figtomography}, we find a mean experimental fidelity $F = 0.996(3)$ for both initial states $\rho_S^{1,2}$.
Throughout our manuscript, we neglect the dynamical variation $F(t)$ in order to detect $\rho_S(t)$ independently from $\rho_E(t)$ and measure trace distances $D(t)$ with uncertainties given by $\delta D(t)$ (see above).

\section{Numerical Simulations}
\label{secNumCalc}

To assess systematic effects in the quantification of quantum memory effects associated with QPN and sampling, we implement numerical simulations of the Hamiltonian $H$, cf.\,\cite{clos_time-resolved_2016}. For tractability, we introduce a Fock-state cutoff $n_c = 20$, which includes more than 99.9988\% of the initial state population for the worst case of our experimentally realized $\rho_E$, i.e., $\bar{n}=1.4$. 
We perform calculations with parameters $\omega_z$, $\omega_E$, and $\Omega$, that we extract from independent calibration measurements to match the experimental realization.

To estimate true values for the non-Markovianity $\mathcal{N}_\text{true}$, we calculate time evolutions $D(t)$ with sampling rate $\gamma = 100 \gamma_0$ and $r \rightarrow \infty$. To judge the accuracy of our estimations, we calculate values for $\gamma = 200 \gamma_0$. We ensure for all presented numerical simulations of $\mathcal{N}_\text{true}$ that $\mathcal{N}(200\gamma_0)/\mathcal{N}_\text{true} < 10^{-3}$.
We incorporate QPN in our numerical simulations by following this five-step recipe:
\begin{enumerate}
	\item Simulate true values $\langle\sigma_l\rangle_\text{true}\;(l=x,y,z)$ for both initial states $\rho_S^{1,2}$.
	\item Calculate the QPN-induced uncertainties $\delta \langle \sigma_l \rangle_\text{true}$ according to Eq.\,\eqref{qpnuncertainty} with corresponding measurement repetitions $r$.
	\item Generate random numbers $\langle\sigma_l\rangle$ with a Gaussian probability distribution centered at $\langle \sigma_l \rangle_\text{true}$ with width (1\,s.d.) $\delta \langle \sigma_l \rangle_\text{true}$.
	\item Calculate trace distance $D$ with incorporated QPN from $\langle\sigma_l\rangle$.
	\item Extract $\mathcal{N}$ from these $D$ to yield one realization of the non-Markovianity with incorporated QPN.
\end{enumerate}
In order to obtain time evolutions $D(t)$ with mean QPN impact, we repeat steps one to four for $k$ times and average the resulting $D$. Accordingly, to obtain $\mathcal{N}$ with mean QPN impact, we repeat steps one to five for $k'$ times and average the resulting $\mathcal{N}$. We choose $k=k'=50$ for our simulations and, thereby, fluctuations between different runs with equal parameters remain non-significant.

\clearpage
\section{Systematic study of $\mathcal{B}(\gamma,r)$}

In order to study the impact of QPN and sampling on the measure for non-Markovianity $\mathcal{N}$, we vary the sampling rate $\gamma$ and measurement repetitions $r$, while keeping other parameters fixed. To minimize effects of systematic uncertainties of the latter, we vary $\gamma$ and $r$ retrospectively, by employing variable postselection procedures.

We vary $r$ for a given measurement $\langle \sigma_l(t) \rangle$ with $r_0$ repetitions by random postselection of subensembles with $r$ of the $r_0$ realizations for each $\langle\sigma_l\rangle$. 
Thereby, we generate new results for each data point $\langle \sigma_l(t) \rangle$ with corresponding uncertainties according to Eq.\,\eqref{qpnuncertainty}. 
We extract $D(t)$ and associated $\mathcal{N}$ and average over 100 iterations to obtain mean values $\mathcal{N}(\gamma_0,r)$.

In order to vary $\gamma$, we proceed analogously: 
Measurements of $\langle \sigma_l(t) \rangle$ with sampling rate $\gamma_0$ for interaction durations $t \in \left[0,t_\text{max}\right]$ create $M_0 \equiv \gamma_0t_\text{max}$ data points $D(t)$. 
To vary $\gamma$, we randomly postselect a subensemble with $M$ of the $M_0$ data points, yielding a measurement of the time evolution $D(t)$ with mean sampling rate $\gamma = (M-1)/t_\text{max}$. 
We extract $\mathcal{N}$ and average over 100 iterations to obtain $\mathcal{N}(\gamma,r_0)$.

\begin{figure}[ht]
	\centering
	\includegraphics{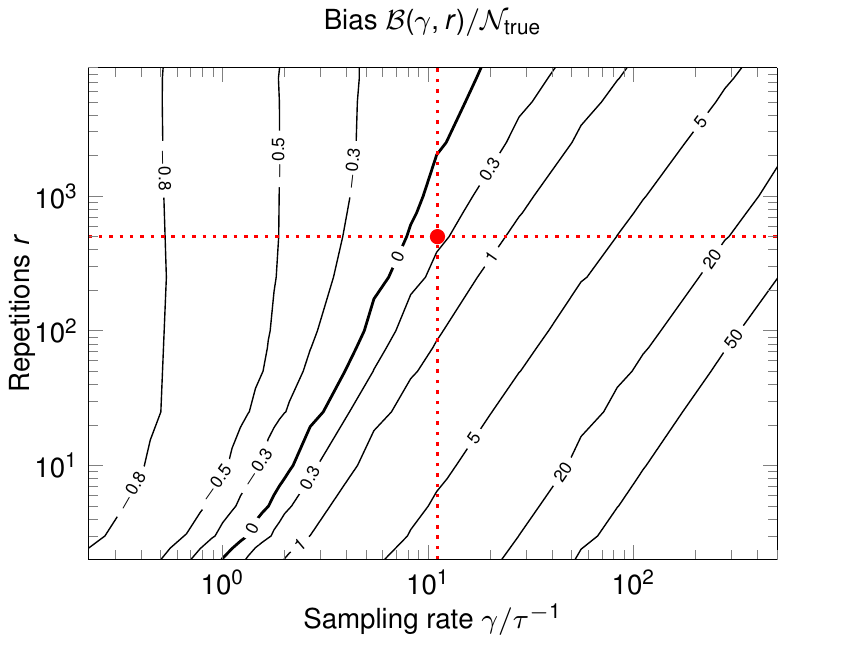}
	\caption{
		$\mathcal{B}(\gamma,r) / \mathcal{N}_\text{true}$ as a function of $\gamma$ and $r$. The dot marks the experimental parameter combination $(\gamma_0,r_0)$, while dotted lines represent the cuts from Fig.\,3.
		\label{figbias}
	}
\end{figure}

The bias $\mathcal{B}$ is an explicit function of $\gamma$ and $r$, i.e., $\mathcal{B} = \mathcal{B}(\gamma,r)=\mathcal{N}(\gamma,r) - \mathcal{N}_\text{true}$. 
For our system, which we operate in a regime where numerical calculations can be performed, we can systematically study $\mathcal{B}$ as a function of $\gamma$ and $r$. 
In Figure\,\ref{figbias} we depict simulated values of $\mathcal{B}(\gamma,r)$ with experimental parameters according to the measurements depicted in Figs.\,2 and 3.  

As depicted, $\mathcal{B}(\gamma,r)$ is a nontrivial function, that diverges for $\gamma \rightarrow \infty$. As $r$ is increased, the divergence flattens, but cannot be eliminated for realistic values of $r$ for our system.
We emphasize that increasing $\gamma$ inevitably results in a growth of $\mathcal{N}$. Thus, any increase of $\gamma$ needs to be accompanied with substantial enhancement of $r$ in order to avoid an increase of $\mathcal{B}$. Moreover, to find optimal settings for $\gamma$ and $r$, 
involved optimization strategies are needed, in particular, if numerical simulations are not available.

\section{Comment on general strategies}

In case of more complex environmental states and interactions, our approach to determine $\mathcal{B}$ based on exact numerical simulations of underlying quantum dynamics may fail. 
In such cases, we suggest to use, e.g., semi-empirical models or Fourier series that are optimized to describe experimental data points $\langle\sigma_l\rangle$. 

For example, semi-empirical models would be optimized based on physical insights that set specific properties of $S$, $E$, and $S$-$E$ interactions. 
Once a model is established, it can be tested to describe experimental data. 
In particular, the variation of $\mathcal{N}(\gamma, r)$ can be generated following the five-step recipe given above, where true values $\langle\sigma_l\rangle_\text{true}\;(l=x,y,z)$ are calculated based on the model. Finally, when data and model calculations are in agreement, the amount of $\mathcal{B}$ can be estimated.

\section{Quantum Probe Applications}

\begin{figure}[htb]
	\centering
	\includegraphics[width=7.6cm]{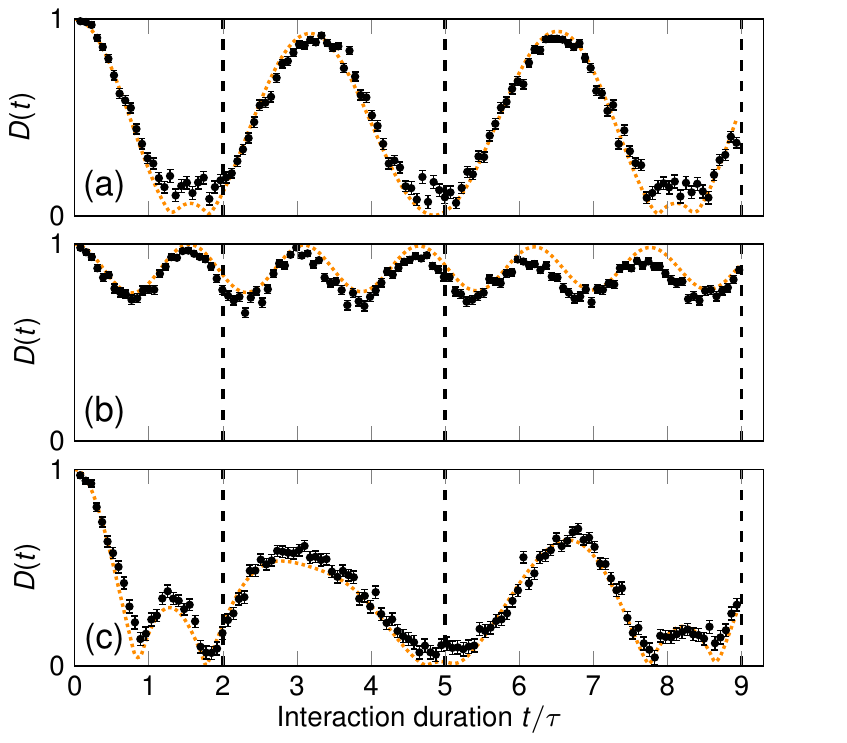}
	\caption{
		Measured $D(t)$, for three representative data points of $\mathcal{N}$ that are shown in Fig.\,4, are compared to true numerical simulations. 
		(a) $D(t)$ for $\omega_z/ \omega_E=1.000(2)$ and $\bar{n} = 0.09(2)$.
		(b) $D(t)$ for $\omega_z / \omega_E=0.900(2)$  and $\bar{n} = 0.09(2)$.
		(c) $D(t)$ for $\omega_z / \omega_E=1.000(2)$ and $\bar{n} = 0.80(2)$.
		Vertical dashed lines depict interaction durations $t_\text{max} = \{2,5,9\}\tau$.
		\label{figsuppFig4}
	}
\end{figure}

We understand Figs.\,4(a) and (b) with the following considerations. 
When tuning $\omega_z$ nonresonant to $\omega_E$, effective coupling rates increase $\Omega' \propto \sqrt{\Omega^2 + (\omega_z-\omega_E)^2}$, see Figs.\,\ref{figsuppFig4}(a) and (b). Thus, short $t_\text{max}$ may show increased $\mathcal{N}$ for nonresonant $S$-$E$ interactions. However, for longer $t_\text{max}$, line shapes become dominated by the resonant $S$-$E$ interaction, since amplitudes in $D(t)$ are $\propto \Omega^2/\Omega'^2$, i.e., they are largest for $\omega_z=\omega_E$.
Additionally, in Fig.\,\ref{figsuppFig4}(b), we observe the impact of decoherence within the observed interaction durations. By comparing measured and simulated $D(t)$, we extract a decoherence rate $\Gamma_\text{dec} \approx 0.06\tau^{-1}$. Throughout our manuscript, we neglect this effect and consider our total system $S$ + $E$ completely isolated from external baths, cf.\,\cite{clos_time-resolved_2016}. 

Comparing Figures\,\ref{figsuppFig4}(a) and (c) yields insight into the effect of varying initial temperature on $\mathcal{N}$ and observed temperature-induced speed-up of memory effects. While overall amplitudes in $D(t)$ are reduced, the dynamics becomes less trivial and partly faster.  

\end{document}